\documentclass{INTERSPEECH2023}

\usepackage{booktabs} 
\usepackage{multirow}
\usepackage{subfigure}
\usepackage{threeparttable}
\usepackage{svg}

\interspeechcameraready

\begin{document}
\pdfoutput=1
\title{Latent Phrase Matching for Dysarthric Speech}
\name{Colin Lea$^*$, Dianna Yee$^*$, Jaya Narain, Zifang Huang, \\Lauren Tooley, Jeffrey P. Bigham, Leah Findlater}

\address{Apple, USA}

\email{\{colin\_lea, dianna\_yee, jnarain, zhuang7, ltooley, jbigham, lfindlater\}@apple.com}

\maketitle
\def\thefootnote{*}\footnotetext{These authors contributed equally to this work}\def\thefootnote{\arabic{footnote}}

\begin{abstract}

Many consumer speech recognition systems are not tuned for people with speech disabilities, resulting in poor recognition and user experience, especially for severe speech differences. Recent studies have emphasized interest in personalized speech models from people with atypical speech patterns. We propose a query-by-example-based personalized phrase recognition system that is trained using small amounts of speech, is language agnostic, does not assume a traditional pronunciation lexicon, and generalizes well across speech difference severities. 
On an internal dataset collected from 32 people with dysarthria, this approach works regardless of severity and shows a 60\% improvement in recall relative to a commercial speech recognition system. On the public EasyCall dataset of dysarthric speech, our approach improves accuracy by 30.5\%. 
Performance degrades as the number of phrases increases, but consistently outperforms ASR systems when trained with 50 unique phrases.

\end{abstract}

\section{Introduction}

Many people with speech disabilities, such as dysarthria, stuttering, or dysphonia, have shown interest in speech technology, but poor automatic speech recognition (ASR) in consumer products has limited adoption, especially for people with severe speech differences~\cite{projecteuphonia,Young_litreview,stuttering_chi23}.

In this work we focus on dysarthria, which is characterised by unclear speech due to weakness or limited control of articulatory movement that may arise developmentally (e.g., due to cerebral palsy or Down syndrome) or may be acquired (e.g., due to ALS, brain injury, or MS)~\cite{Young_litreview}. 
For people with dysarthria, pronunciations vary person-to-person, where some may not enunciate certain sounds or may have distorted vocalizations relative to other parts of the general population.  Since general-purpose ASR systems commonly assume that words are broken into sub-units (phones or sub-words) that are pronounced similarly across individuals, divergence from those pronunciations tend to significantly reduce accuracy.

Recent efforts on dysarthric speech recognition focused on personalized or tuned ASR models (e.g., \cite{uaspeech_source_domain_data_selection,green2021automatic, tobin2022personalized, Shor_2019,uaspeech_wav2vec_speakeradapt,Tomanek_2021,xiao2021scaling}), which leverage large proprietary or non-commercial datasets of atypical speech (e.g., Project Euphonia~\cite{projecteuphonia}; UASpeech \cite{uaspeech_asr}; AphasiaBank~\cite{aphasiabank}). 
In this work, we take a more pragmatic approach that does not require vast quantities of data, yet enables people with severe speech differences to train phrase recognition models for applications where only a constrained set of phrases is needed. 

Our approach, Latent Phrase Matching (LPM), builds on recent efforts on personalized keyword spotting (KWS) ~\cite{Lopez_kws_overview, Yang_kws_personalized_multitask, Mazumder_few_shot_KWS_any_language} and query-by-example (QBE) ~\cite{DONUT_ctc_query_by_example_KWS,Hazen_phonetic_query_by_example_KWS,Rodriguez_phones_query_by_example_KWS, Chen_query_by_example_LSTM} to recognize an arbitrary number of phrases of arbitrary duration.
Using a small set of speech samples, we create a model per-phrase using latent embeddings from a large-scale keyword spotter that is agnostic to phonetic content or prosody.
At run-time, new vocalizations are compared to enrolled phrases using a dynamic time warping metric that is flexible to variations in speaking rate and style. 

We assess our approach, along with a commercial ASR system that is tuned for the general public, on two datasets. First, we use an internal collection of English speech from 32 people with dysarthria, collected across multiple sessions and times of day with the goal of capturing variability in an individual's speech characteristics. 
Second, we use the EasyCall~\cite{easycall} dataset of Italian speech from 31 people with dysarthria. 
In contrast to other public dysarthria datasets (e.g., Torgo~\cite{torgo}), both evaluation datasets include recordings from each individual of tens of unique phrases at least five times each. 
Our contributions are as follows. 
We validate existing research on atypical speech (e.g., \cite{projecteuphonia}) by demonstrating large differences in ASR performance for people with varying speech disabilities.
We propose a speech embedding model that encodes dysarthric speech patterns better than existing models and pair it with a query-by-example framework for phrase recognition. 
We validate the approach on two datasets totalling 63 people with dysarthria across two languages and demonstrate that phrase recognition performance is much higher for people with severe dysarthria than traditional ASR systems.
Furthermore, we discuss additional design decisions to consider when collecting dysarthric speech data and developing and validating dysarthric phrase recognition systems.

\begin{figure*}[t!]
    \hfill
    \subfigure[Phrase enrollment and run-time inference using Latent Phrase Matching.] {\includegraphics[width=0.5 \paperwidth]{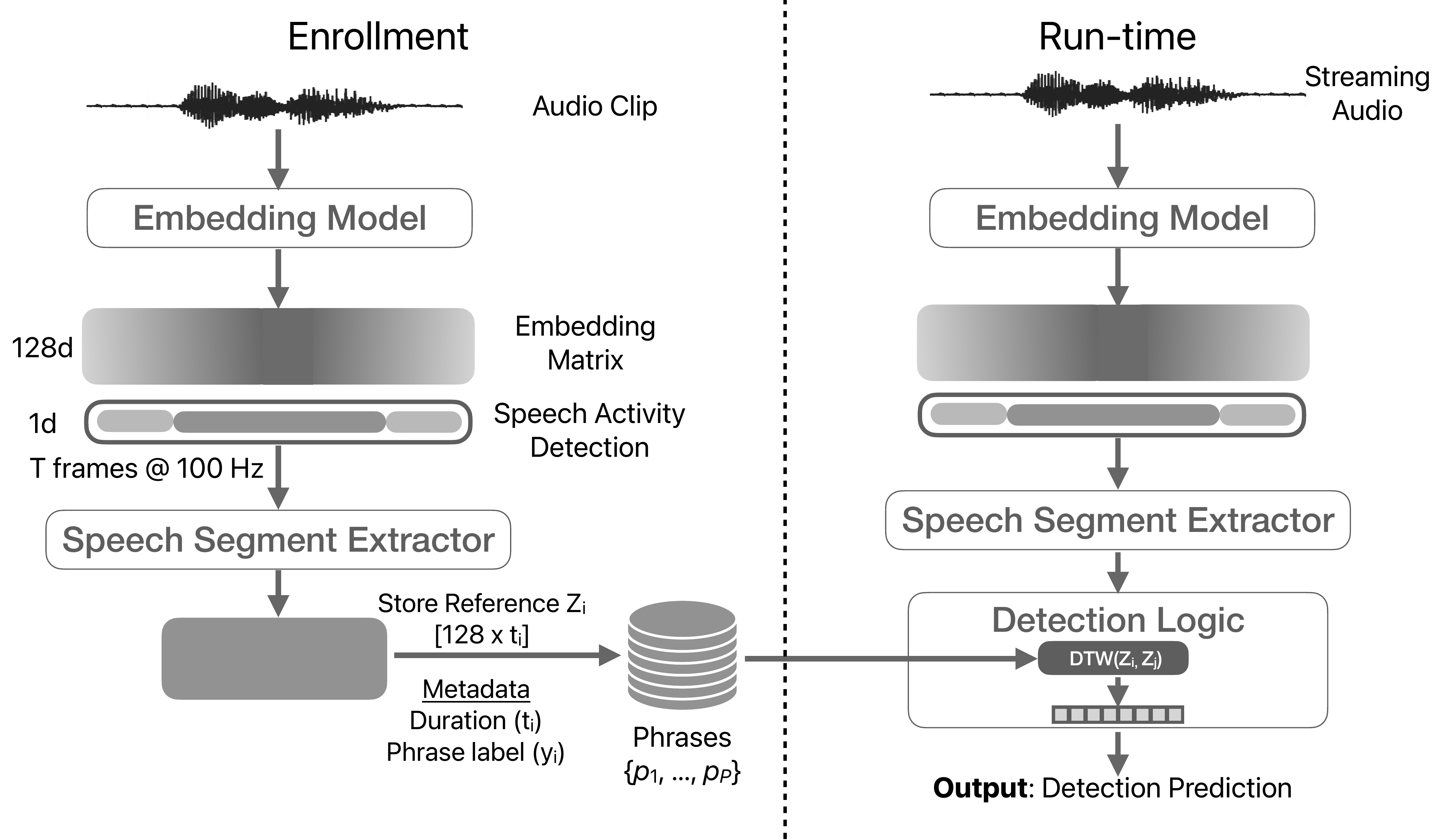}}
    \hfill
    \subfigure[Speech Embedding Model] {\includegraphics[width=0.22 \paperwidth]{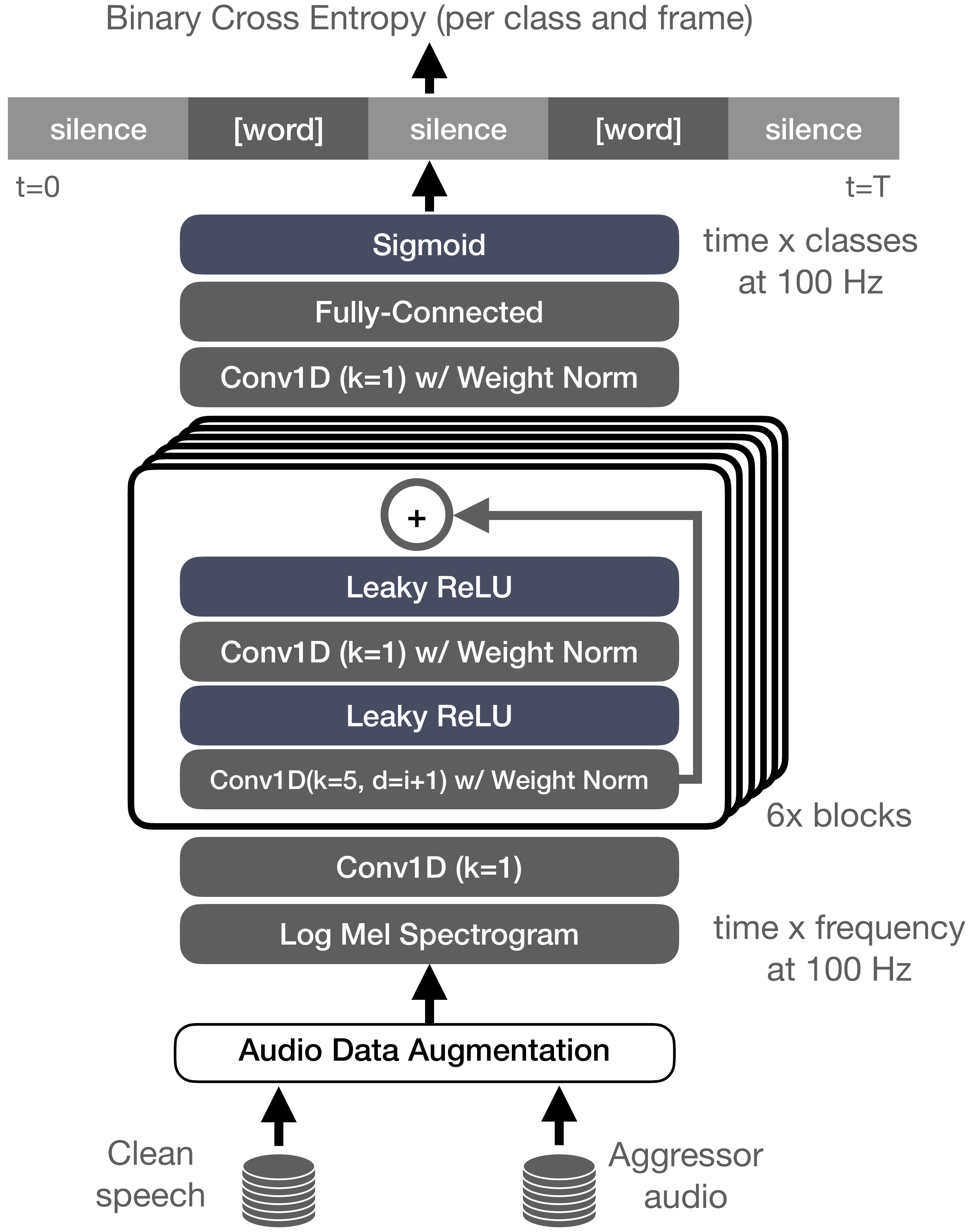} }
    \hfill
    \caption{(a) Overview of phrase enrollment and inference using a Query-by-Example approach. (b) Architecture and training design details of the speech embedding model. }
    \label{fig:approaches}
\end{figure*}

\section{Methods}

Our LPM method uses a QBE approach that consists of a speech embedding model that is optimized for supervised word recognition to create a model per-phrase, and a similarity metric to compare phrases using dynamic time warping (DTW)~\cite{senin2008dynamic} on latent feature representations.
In comparison to~\cite{Amin_2008} who computes DTW directly on mel-frequency cepstrum coefficients (MFCCs), our approach, like more recent approaches~\cite{QbyE-MLPMixer}, uses neural networks for feature learning, such that generally relevant but complex speech patterns can be extracted for open-vocabulary keyword modeling. 
Instaed of relying on phone-based approaches as in~\cite{Hazen_phonetic_query_by_example_KWS, Rodriguez_phones_query_by_example_KWS},
our speech embedding model is designed to encode word- and phrase-like attributes. This is because dysarthric speech characteristics may vary with time and can lead to inconsistency in the predicted phones of a given phrase and subsequently poor phrase recognition performance.
Additionally, our approach does not require any training of model parameters when a new phrase is onboarded, unlike the supervised approaches proposed in~\cite{Yang_kws_personalized_multitask, DONUT_ctc_query_by_example_KWS}.

\subsection{Latent Phrase Matching (LPM)}
At training time, a user enrolls a phrase by speaking each phrase at least twice and a model is built per-phrase using a speech embedding model. 
The model for each phrase is parameterized by tuple $p_i = \big( Z_i, y_i, \tau_i \big)$ for index $i \in \{1...P\}$ where $Z_i$ is a latent phrase representation, $y_i$ is a phrase label, $\tau_i$ is a within-phrase detection threshold, and $P$ is the number of enrolled phrases. 

At run-time, candidate phrase segments are extracted from audio and compared to each enrolled phrase using DTW. 
If a spoken phrase is sufficiently similar to an enrolled phrase, that phrase is \textit{detected}, otherwise it is classified as an out of domain utterance. 
The enrollment and inference algorithms are depicted in Figure~\ref{fig:approaches}.

\subsubsection{Phrase Representation} 
We posit that a phrase representation for dysarthric speech should have the following properties: 
(1) it encodes local word-like structure that supports phrases constructed from an unbounded vocabulary, but embeddings for two people saying the same word need not be similar; 
(2) it encodes global structure such as sequencing of word-like information, but acknowledge that the cadence or vocal quality may vary throughout the day or as speech characteristics change; 
(3) the embeddings are compact such that the model does not overfit to spurious correlations from a small number of training examples;
(4) lastly, the speech embeddings are robust to background noise or acoustic changes in the environment. 

We satisfy (1) and (3) by constructing a sufficiently large vocabulary keyword detection model and using low-dimensional word-like intermediate embeddings. 
We satisfy (2) by computing embeddings at a frame rate of 100 Hz and storing as templates to use with DTW to compute phrase similarities that are invariant to signal shifting and scaling in time.
For (4), our keyword spotting model is trained by mixing copious amounts of background noise randomly with speech such that the learned speech representations are implicitly denoised.

Our keyword model in Figure~\ref{fig:approaches}(b) is a ResNet-style temporal convolutional network designed using many recent best practices for large-scale CNN training~\cite{wightman2021resnet} and is trained on subsets of the Multilingual Spoken Words Corpus (MSWC) \cite{mazumder2021multilingual}. 
The input to this model is 64-dimensional log melspectrograms with a frame rate of 100 Hz. The output is per-frame keyword predictions at 100 Hz for 300 common English words plus an extra label for speech activity detection.\footnote{300 words chosen from MSWC that had at least 1000 instances and various lengths from 3 to 11 letters. For early experiments we trained models with different languages, including German and Italian but results were similar to English-only regardless of the testing language. } 
The architecture consists of six blocks of dilated convolutions with the pattern \texttt{Conv1D(k=5, d=i+1), LeakyReLU, Conv1D(k=1), LeakyReLU} for the $i$-th block where \texttt{k} is kernel size,  \texttt{d} is dilation rate, and the receptive field is 850 ms.
Each block has a residual connection, weight norm~\cite{weightnorm} in convolutional layers, and dropout during training. 
Fully connected layers are used to project the latent features to the desired output dimensions  and the final block has a sigmoid activation which is compatible with the Binary Cross Entropy (BCE) loss function. 
We use a BCE loss per frame where at most one of the possible words in our vocabulary is active.
For training purposes, we concatenate MSWC clips together so each batch is of the form \texttt{[sil] [word] [sil] [word] [sil] ...}, where start and stop times for each word are estimated using the MSWC annotations.
To improve robustness, we randomly added background sounds to audio clips. 

For our phrase representations we use intermediate embeddings,\footnote{The specific embedding layer was determined based on early experimentation using EasyCall and was not changed for use with our internal dataset.} where given an audio clip with $t$ spectral frames, the output is an embedding matrix $Z_i$ also with length $t$ frames and with height $f = 128$.
The speech activity detector label is used to truncate silence at the beginning and end of a clip. The speech activity detector is applied at both enrollment and evaluation such that periods of predicted speech absence are not considered in the similarity computation.

\subsubsection{Phrase Similarity} 
 
During enrollment of the $i$-th phrase within $\mathcal{P}$, we compute the pair-wise similarity between all phrases using DTW to determine a within-phrase threshold $\tau_i$ by
\begin{equation}
    \tau_i = \alpha \cdot \big( \max_{\substack{ j \in \{1...P\} \\ y_i = y_j,  i \neq j}} \, \mathrm{DTW}(Z_i, Z_j) \big),
\end{equation}
where $\alpha$ is a tuneable constant (e.g., $\alpha = 1.25$). 
The purpose of $\tau_i$ is to reject phrases that are not part of the enrolled set (e.g., read or conversational speech). When $\alpha=\infty$  this detector becomes a phrase classification model. The choice of $\alpha$ is determined by the desired trade-off between recall and precision. 

At run-time, if the computed DTW score is lower than any within-phrase threshold, then $y_j$ is the detected phrase, where $j$ indicates the $j$-th phrase with the lowest score, otherwise no phrase was detected. 

\section{Data}
\subsection{Internal Collection}
To validate our approach, we collected audio recordings from 32 participants with varying severity of dysarthria who identified as having a motor-speech disability. The data collection spanned five sessions across multiple days to observe how speech characteristics differed over time. In each session, a participant repeated phrases five times each with a microphone placed firstly close to their body and then farther on a table in front of them.
In total, there were 50 unique commonly spoken phrases like ``Scroll up" or ``Go to the home screen". Additionally, participants read paragraph-long passages and spoke about two topics freely. A Speech-Language Pathologist (SLP) listened to and graded the data to assess intelligibility and severity according to the Dysarthria Rating of Severity scale~\cite{yorkston1999management} which ranges from 1-5 (1: no detectable speech disorder, 2: obvious  speech disorder with intelligible speech, 3: reduction in intelligibility 4: severe dysarthria with significant reduction in intelligibility, and 5: minimal to no speech production). Our dataset only contains participants who have been graded 2-4, so in this work, we refer to 2 as mild, 3 as moderate, and 4 as severe. 

\subsection{EasyCall Dataset}
We also use the EasyCall dataset~\cite{easycall} which contains recordings from 31 participants with dysathria and 25 without dysathria. Each participant vocalized in Italian 20 phrases and 46 words repeated over 2 to 8 sessions. 
For each participant, we only evaluated on phrases that had at least three repetitions across sessions. 
We found many clips were incorrectly truncated or had significant background speech from a non-participant and excluded these from our experiments as part of a QA process; in total we used 18411 of 21386 (86\%) audio clips.\footnote{Please email the corresponding authors for specific filenames.}

\section{Results}
\begin{figure}
    \centering
    \includegraphics[width=0.8\linewidth]{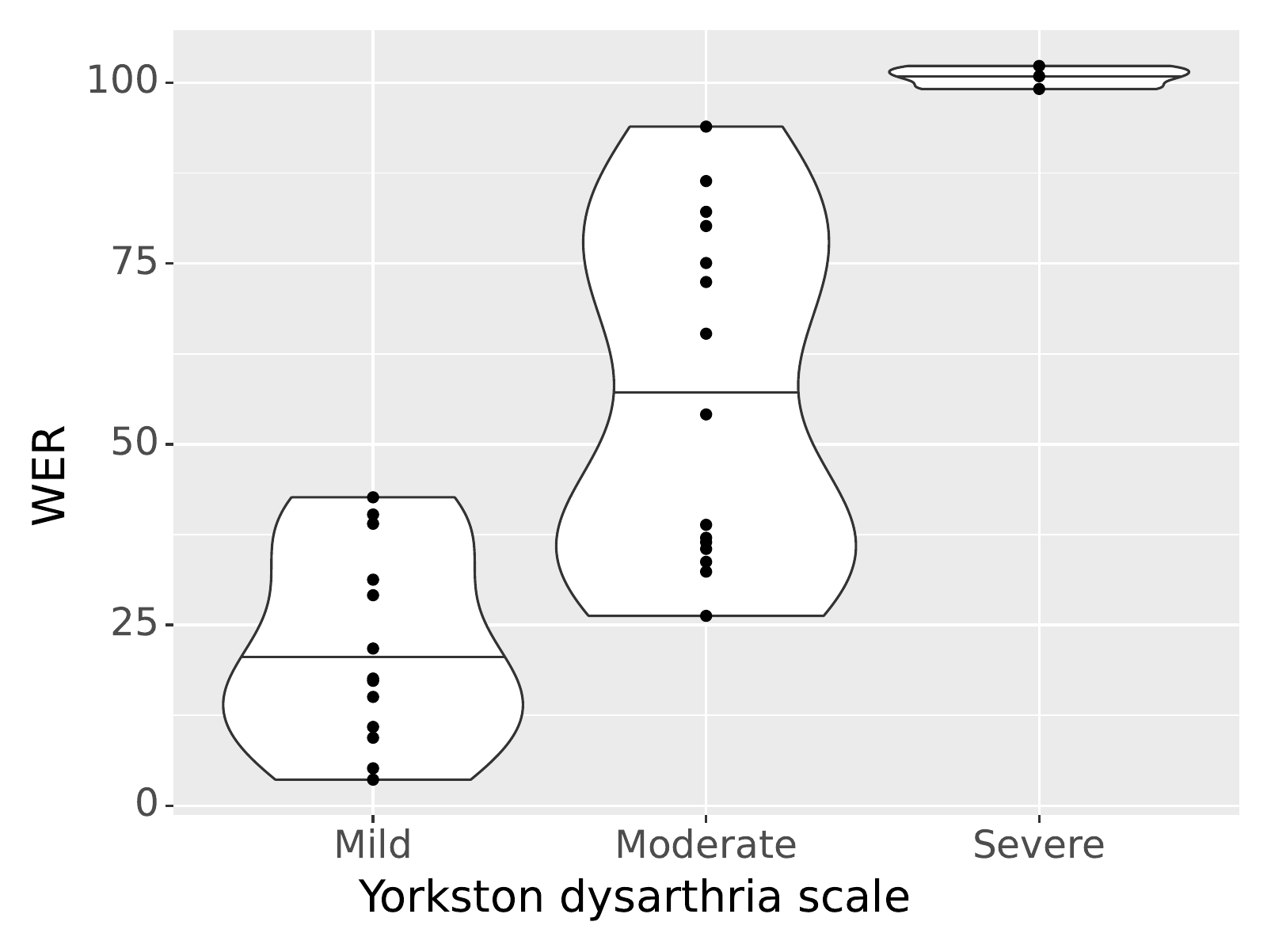}
    \caption{Per-participant Word Error Rates (WERs) of a consumer ASR model on our internal dataset. Severity is graded by an SLP using the Yorkston dysarthria scale. Significant Spearman's rank correlation ($r(30)=0.76$, $p<.001$) support ASR performance degrades as dysarthria severity increases.}
    \label{fig:BaselineWERYorkshireSeverity}
\end{figure}

In our experiments, we use the Apple Speech Framework~\cite{SpeechFramework} as a baseline ASR system and define $\mathcal{P}$ as all the phrases within our dataset. 
The WER of the $\mathrm{ASR}$ on our internal dataset is significantly correlated with dysarthria severity as seen in
Figure \ref{fig:BaselineWERYorkshireSeverity}, and verifies that ASR does not fully support moderate-severe dysarthria. The average WERs on EasyCall phrases is 53.4\% for people with dysarthria and and 8.7\% for the control group.

\subsection{Phrase Recognition Benchmark}
\label{subsec:phrase_rec_benchmark}
For each phrase within $\mathcal{P}$, a model is trained with utterances recorded within the same recording session and evaluated on utterances exclusive to the training session. We evaluate for both in-domain and out-of-domain scenarios. In-domain experiments consider only spoken utterances within $\mathcal{P}$ and evaluate phrase recognition using canonical classification metrics like recall and precision \cite{intro_nlp_Eisenstein}. 
We report the in-domain performance metrics in Table \ref{tab:phrase_recog_eval} based on 5 repeated trials. We observe that phrase recognition performance generally decreases with severity, however the disparity in performance is much higher for the $\mathrm{ASR}$. LPM has significantly better recognition performance for severe and moderate speech dysarthria.  We also evaluate for false detection rate (FDR) where FDR is defined as the frequency aggressor speech is falsely detected as one of the phrases within $\mathcal{P}$. The aggressor speech used comes from the read passages and are exclusive to $\mathcal{P}$. The FDRs for the $\mathrm{ASR}$ and LPM are respectively 0.0 and 0.34$\pm$0.22. $\mathrm{ASR}$ has a low FDR since it is able to detect a larger vocabulary of words to discern between aggressor speech and enrolled phrases.  

To quantify variability in phrase recognition performance arising from microphone placement and utterances recorded on the same day as training versus a different day, we report on the average per-participant accuracies for phrases recorded near or farther during the same session as training or on sessions from other days in Table \ref{tab:phrase_recog_eval_session_bar}. We observe that performance variability arose due to microphone placement and when recordings were made on a different day. Accuracies are typically the highest for conditions most similar to the training session. For moderate to severe, we observed performance degrades when the recording was made with the microphone placed farther or on a different day than the training session.

\begin{table}[h]
    \centering
    \caption{Mean and standard deviation of per-participant phrase detection metrics of baseline ASR system ($\mathrm{ASR}$) and proposed LPM for 50 enrolled phrases.} 
    \label{tab:phrase_recog_eval}
    \begin{tabular}{@{}p{0.95cm}p{0.95cm}lll@{}}
        \toprule
        {Models}& {Severity} & Accuracy & Precision & Recall   \\
        \midrule
        \multirow{3}{*}[0em]{$\mathrm{ASR}$} & mild & 0.78$\pm$0.13  & 0.92$\pm$0.07   & 0.73$\pm$0.14\\
         & moderate & 0.44$\pm$0.23  & 0.67$\pm$0.26  & 0.40$\pm$0.23 \\
         & severe & 0.02$\pm$0.01  & 0.07$\pm$0.05   & 0.02$\pm$0.01 \\
         & all & 0.54$\pm$0.30  & 0.71$\pm$0.31  &   0.50$\pm$0.29 \\
         \midrule
        \multirow{3}{*}[0em]{LPM}& mild & 0.84$\pm$0.04  & 0.87$\pm$0.04  & 0.84$\pm$0.04 \\
         & moderate & 0.78$\pm$0.15  & 0.80$\pm$0.14   & 0.78$\pm$0.14 \\
         & severe & 0.70$\pm$0.16  & 0.73$\pm$0.18  & 0.70$\pm$0.17 \\
         & all & 0.80$\pm$0.12  & 0.82$\pm$0.12  & 0.80$\pm$0.12\\
        \bottomrule
    \end{tabular}
\end{table}

\begin{table}[h]
    \centering
    \caption{Mean and standard deviation of per-participant phrase detection accuracy conditioned on utterances recorded within the same or exclusive to training sessions.}
    \label{tab:phrase_recog_eval_session_bar}
    \begin{tabular}{@{}p{0.95cm}p{0.95cm}lll@{}}
        \toprule
 
        {Models}& {Severity} & Same (near) & Same (far)   & Other \\
        \midrule
        \multirow{3}{*}[0em]{$\mathrm{ASR}$} & mild & 0.80$\pm$0.18  & 0.81$\pm$0.18   & 0.80$\pm$0.18\\
         & moderate & 0.52$\pm$0.25  & 0.45$\pm$0.25  & 0.46$\pm$0.25 \\
         & severe & 0.00$\pm$0.01  & 0.01$\pm$0.01  & 0.02$\pm$0.01 \\
         \midrule
        \multirow{3}{*}[0em]{LPM}& mild & 0.76$\pm$0.29  & 0.76$\pm$0.19  & 0.75$\pm$0.17 \\
         & moderate & 0.88$\pm$0.09  & 0.79$\pm$0.23   & 0.81$\pm$0.13 \\
         & severe & 0.94$\pm$0.10  & 0.80$\pm$0.22  & 0.68$\pm$0.25 \\
        \bottomrule
    \end{tabular}
\end{table}

\begin{table}[h]
    \centering
\caption{ Phrase recognition accuracy on the EasyCall dataset with a baseline ASR system and variants of LPM using spectral features or KWS embeddings within the phrase representation.}
    \label{tab:easycall}
\begin{tabular}{@{}l c c c}    
        \toprule
Data Subset         & $\mathrm{ASR}$ & LPM$_\mathrm{Spec}$ & LPM$_\mathrm{KWS}$ \\
\toprule

Control (word)      & 0.85$\pm$0.04 & 0.79$\pm$0.15    & 0.84$\pm$0.12    \\
Control (phrase)    & 0.87$\pm$0.07 & 0.90$\pm0.11$       & 0.95$\pm0.09$    \\
Dysarthria (word)    & 0.55$\pm$0.20 & 0.61$\pm$0.21    &  0.65$\pm$0.20   \\
Dysarthria (phrase) & 0.63$\pm$0.26 & 0.72$\pm0.21$        & 0.82$\pm0.16$   \\
\bottomrule
\end{tabular}
\end{table}

\subsection{Ablation studies}

\subsubsection{EasyCall Experiments}
We benchmark our approach on the EasyCall dataset against the ASR baseline where a phrase is considered to be correct if the ASR can verbatim recognize the onboarded phrase.
Results are shown in Table~\ref{tab:easycall}.
In addition, we compare against a baseline using melspectrogram features as the phrase representation. 
In general, longer phrases, which tend to have a richer latent structure, perform better across approaches. 
The DTW-based approaches perform significantly better than the ASR baseline, especially for dysathric phrases where LPM accuracy is 30.5\% higher than ASR. 
Based on qualitative analysis, the Spectral DTW baseline appears to overfit to environment conditions and variation in articulation much more than the KWS-DTW approach. 
We also compared our embedding model to a Conformer-based ASR encoder~\cite{Gulati_2020} using ESPNet~\cite{watanabe2018espnet} that was trained on Librispeech, which yields very similar results to our KWS embeddings, albeit at much higher computational cost.\footnote{Results not shown due to space limitations.}

\subsubsection{Varying number of phrases}
\begin{figure}
    \centering
    \includegraphics[width=\linewidth]{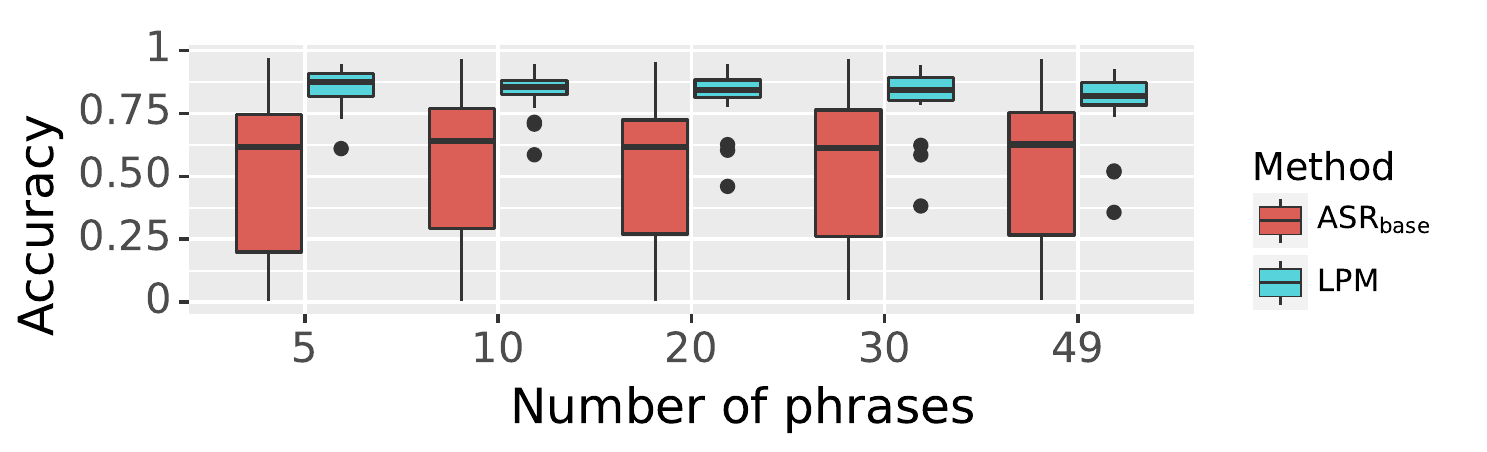}
    \caption{Distribution of average per-participant accuracy of the baseline ASR system and LPM for varying number of enrolled phrases.}
    \label{fig:nCx_ablation}
\end{figure}

To investigate the sensitivity of LPM to the number of unique enrolled phrases, denoted by $n$, the experiment in Section \ref{subsec:phrase_rec_benchmark} was repeated with various values of $n \in$ [5, 10, 20, 30, 49].
For $n$=5 we ran 10 trials and for other values of $n$ we ran 5 trials. The distribution of the averaged per-participant metrics are plotted in Figure \ref{fig:nCx_ablation}.
The $\mathrm{ASR}$ does not rely on the training phrases and thus is independent of $n$. For moderate and severe, LPM performance degrades as $n$ increases. This is represented by the points outside of the interquartile ranges in Figure \ref{fig:nCx_ablation}.
Future work should look at improving robustness for participants with these speech patterns. 

\subsubsection{Varying Signal-to-Noise Ratios}
To investigate the influence of common household noises on detection performance, 
we contrived scenarios where the phrases for evaluation were augmented with background sounds at  different signal-to-noise ratios (SNR). The recognition performances are shown in Table \ref{tab:SNR_ablation} where we observe degradation of recognition performance as SNR decreases for all approaches. The degradation also seems to be magnified with dysarthria severity. Regardless of the level of environmental noise, we observe that LPM exhibits less model bias than the $\mathrm{ASR}$ and could conduce to a more inclusive phrase recognition system than an ASR-based approach.

\begin{table}[tbh]
    \centering
    \caption{Mean and standard deviation of Precision of the baseline ASR and LPM in presence of additive background noise contrived synthetically with recorded household sounds. }
  \label{tab:SNR_ablation}
    \begin{tabular}{@{}lllll@{}}
        \toprule
        Models & Severity &  5 dB &  20 dB &  $\Delta$  \\
        \midrule
        \multirow{3}{*}[0em]{$\mathrm{ASR}$} & mild & 0.87$\pm$0.12  & 0.95$\pm$0.04 & 0.08 \\
        & moderate & 0.42$\pm$0.30  & 0.59$\pm$0.31 & 0.17  \\
        & severe & 0.04$\pm$0.02  & 0.04$\pm$0.02 & 0.00  \\
        \midrule
        \multirow{3}{*}[0em]{LPM} & mild & 0.72$\pm$0.07 & 0.85$\pm$0.04 & 0.13  \\
        & moderate & 0.64$\pm$0.06 & 0.79$\pm$0.07 & 0.15 \\
        & severe & 0.57$\pm$0.10 & 0.73$\pm$0.07 & 0.16  \\
        \bottomrule
    \end{tabular}
\end{table}

\section{Conclusions}
In this work we propose a personalized phrase recognition algorithm using query by example on latent representations of speech. We conducted a novel data collection that accounts for speech variability for people with varying severities of dysarthria and used the collected data to benchmark our approach with a state-of-the-art ASR system. Experimental results show evidence that our approach exhibits less model bias than using ASR for phrase recognition and potentially enables people with moderate and severe dysarthria to use speech technology. Additionally, we conducted several ablation studies to quantify the sensitivity of performance due to the design choice of speech representations, level of environmental noise, and number of phrases supported for phrase recognition.

\bibliographystyle{IEEEtran}
\bibliography{citations}
\end{document}